\documentclass[preprint2]{aastex}

\slugcomment{To be submitted to \emph{Astrophysical Journal}}

\begin{document}

\title{On the very high energy spectrum of the Crab pulsar.}

\author{Chkheidze N.\thanks{E-mail:
nino.chkheidze@iliauni.edu.ge}, Machabeli G. and Osmanov Z.}
\affil{Center for theoretical Astrophysics, ITP, Ilia State
University, Kazbegi ave. 2$^a$, 0160, Tbilisi, Georgia}

\begin{abstract}
In the present paper we construct a self-consistent theory,
interpreting the observations of the MAGIC Cherenkov Telescope of
the very high energy (VHE) pulsed emission from the Crab pulsar. In
particular, on the basis of the Vlasov's kinetic equation we study
the process of the quasi-linear diffusion(QLD) developed by means of
the cyclotron instability. This mechanism provides simultaneous
generation of low (radio) and VHE (0.01-25GeV) emission on the light
cylinder scales, in one location of the pulsar magnetosphere. A
different approach of the synchrotron theory is considered, giving
the spectral index of VHE emission ($\beta=2$) and the
exponential cutoff energy ($23$GeV) in a good agreement with the
observational data.
\end{abstract}


\section{Introduction}

The recent observations of the MAGIC Cherenkov telescope
\citep{magic}, reveal several characteristic features of VHE
emission of the Crab pulsar. In particular, pulsed $\gamma$-rays
above $25$GeV is detected showing a relatively high energy cut-off,
which indicates that emission happens far out in the magnetosphere
\citep{magic}. In general, for explaining $\gamma$-ray production in
the pulsar magnetosphere two different models are applied: the
so-called 'polar cap' and 'outer gap' models. In the former it is
assumed that the VHE emission is generated at the polar cap (e. g.
\citet{dh} ), but it can not coincide in phase with the entire radio
emission. And in the outer gap models the generation of the VHE
radiation happens in the outer gap region (e. g. \citet{RY}), but to
our knowledge there is no mechanism of generation of radio emission.
The pulsar emission model that underlies our work principally
differs from polar cap and outer gap models. To explain the origin
of the very high energies observed by the MAGIC Cherenkov Telescope
\citep{magic}, we rely on the pulsar emission model first developed
by \citet{machus1} and \citet{lomin}. According to these works, in
the electron-positron plasma of a pulsar magnetosphere the low
frequency cyclotron modes, on the quasi-linear evolution stage
create conditions for generation of the high energy synchrotron
emission. A special interest deserves the coincidence of signals
from different frequency bands ranging from radio to $X$-rays
\citep{manch}. Investigations of last decade have shown that the
aforementioned coincidence takes places in the very high energy
domain ($0.01$MeV-$25$GeV) as well \citep{magic}. In the framework
of the present paper, generation of low and high frequency waves is
a simultaneous process and it takes place in one location of the
magnetosphere, which explains the observed pulse phase coincidence
of the low and VHE signals. Consequently, we suppose that generation
of phase-aligned signals from different frequency bands is a
simultaneous process and takes place in one location of the pulsar
magnetosphere. This consideration, automatically excludes the
inverse Compton scattering and the curvature radiation mechanisms
respectively, which are not localized \citep{difus,difus1}. It is
worth noting that coincidence of pulse phases might be achieved by
means of caustic effects \citep{mo,RY,dyks}.

It is well known that close to the pulsar surface due to very strong
magnetic fields, magnetospheric particles emit efficiently and the
corresponding cooling timescale is short compared to the typical
kinematic timescales of particles. Therefore, transversal energy
loss becomes extremely efficient, consequently electrons and protons
loose their perpendicular momenta and very rapidly transit to their
ground Landau states and the distribution function becomes one
dimensional. This means that one needs a certain mechanism, leading
to the creation of the pith angles restoring the synchrotron
radiation. The main mechanism of wave generation in plasmas of the
pulsar magnetosphere is the cyclotron instability \citep{kmm}.
During the quasi-linear stage of the instability, a diffusion of
particles arises along and across the magnetic field lines.
Therefore, the resonant electrons acquire transverse momenta and, as
a result start to radiate in the synchrotron regime.

Recently we applied the method of the quasi-linear diffusion to the
Crab pulsar to explain the VHE emission observed by the MAGIC
Cherenkov Telescope \citep{magic}. In \citep{difus} we found that on
the light cylinder (a hypothetical zone, where the linear velocity
of rigid rotation exactly equals the speed of light) lengthscales
the cyclotron modes are generated, provoking the re-creation of the
pitch angles and the subsequent synchrotron radiation in the VHE
($>25$GeV) domain. The quasi-linear diffusion guarantees the
observationally evident fact of coincidence of signals in the low
and the VHE bands \citep{difus}, meaning the simultaneous generation
of these radiation domains in one location of the pulsar
magnetosphere. As it has been shown, neither the curvature radiation
nor the inverse Compton scattering may provide the above mentioned
coincidence. This particular problem was considered in
\citep{difus1}, where analyzing the inverse Compton scattering, it
has been demonstrated that for reasonable physical parameters even
very energetic electrons are unable to produce the photon energies
of the order of $25$GeV. Studying the curvature radiation, we have
found that the curvature drift instability \citep{mnras,forcefree}
efficiently rectifies the magnetic field lines making the role of
the curvature emission process negligible \citep{difus1}.

According to the theory of the synchrotron emission
\citep{bekefi,ginz} the spectral index is usually less than $1$ (in
most of the cases it equals $1/2$), although it is observationally
evident that for VHE pulsed radiation from the Crab pulsar the
corresponding value is of the order of $2$ \citep{magic}. In the
standard theory of the synchrotron radiation, unlike the present
model, it is assumed that along the line of sight the magnetic field
is chaotic, leading to the broad interval (from $0$ to $\pi$) of the
pitch angles. Contrary to this scenario, as we have already
outlined, in the pulsar magnetospheres the magnetic field is very
strong and pitch angles rapidly vanish. The present model provides
all necessary conditions for re-creation of the pitch angles,
consequently restricting their values. In the framework of the paper
we study the spectral index, which, for reasonable magnetospheric
parameters is of the order of $2$ for the high energy domain.

The paper is organized as follows. In Section 2 we consider the
emission model, in Sect. 3 we study the synchrotron radiation
spectrum, in Sect. 4 we discuss our results and in Sect. 5 we
summarize them.



\section{Emission model}
\label{sec:consid}
According to the works of \cite{stur} and \cite{tadem} due to the
cascade processes of the pair creation a pulsar's magnetosphere is
filled by electron-positron plasma with  an anisotropic
one-dimensional distribution function (see Fig. 1 from
\cite{arons}) and consists of the following components: the bulk
of plasma with an average Lorentz-factor $\gamma\sim\gamma_{p}$, a
tail - $\gamma_{t}$, and the primary beam with
$\gamma\sim\gamma_{b}$. While considering the eigen-modes of
electron-positron plasma having small inclination angles with
respect to the magnetic field, one has three branches, two of which
are mixed longitudinal-transversal waves ($lt_{1,2}$). The high
frequency branch on the diagram $\omega(k_{_\parallel})$ begins with
the Langmuir frequency and for longitudinal waves ($k_{\perp}=0$),
$lt_1$ reduces to the pure longitudinal Langmuir mode. The low
frequency branch, $lt_2$, is similar to the Alfv\'en wave. The third
$t$ mode, is the pure transversal wave, the electric component of
which ${\bf E^t}$ is perpendicular to the plane  of the wave vector,
and the magnetic field, $({\bf k,B_0})$. The vector of the electric
field, ${\bf E^{lt_1,lt_2}}$ is located in the plane $({\bf
k,B_0})$. When $k_{\perp}=0$, the $t$-mode merges with the $lt$
waves and the corresponding spectra is given by \citep{kmm}
\begin{equation}\label{1}
\omega_t \approx kc\left(1-\delta\right),\;\;\;\;\; \delta =
\frac{\omega_p^2}{4\omega_B^2\gamma_p^3}
\end{equation}
where $k$ is the modulus of the wave vector, $c$ is the speed of
light, $\omega_p \equiv \sqrt{4\pi n_pe^2/m}$ is the plasma
frequency, $e$ and $m$ are the electron's charge and the rest mass,
respectively, $n_p$ is the plasma density, $\omega_B\equiv eB/mc$ is
the cyclotron frequency and $B$ is the magnetic field induction.

The distribution function is one-dimensional and anisotropic and
plasma becomes unstable, which can lead to excitation of the
aforementioned waves. The beam particles undergo drifting
perpendicularly to the magnetic field due to the curvature, $\rho$,
of the field lines. The corresponding drift velocity is given by
\begin{equation}\label{2}
u_x\equiv \frac{cV_{_{_{\|}}}\gamma_{res}}{\rho\omega_B}
\end{equation}
where $V_{_{\|}}$ is the component of velocity along the magnetic
field lines and $\gamma_{res}$ is the Lorentz factor of the resonant
particles. Both of these factors (the one-dimensionality of the
distribution function and the drift of particles) might cause
generation of eigen modes in the electron-positron plasma if the
following resonance condition is satisfied \citep{kmm}
\begin{equation}\label{3}
\omega - k_{_{\|}}V_{_{\|}}-k_xu_x+\frac{s\omega_B}{\gamma_{res}} =
0,
\end{equation}
where $k_x$ is the wave vector's component along the drift and $s =
0,\pm 1, \pm 2,...$

For $s = 0$ one has a hollow cone of the modified Cherenkov
radiation \citep{kmm,lmb,smmk}.  For the Crab pulsar one has the
core emission, being a result of the anomalous Doppler effect ($s =
-1$). Our aim is to interpret the results of the MAGIC Cherenkov
Telescope \citep{magic} therefore, in the present paper we consider
the aforementioned resonance condition $s=-1$.

During the generation of $t$ or $lt$ modes by resonant particles,
one also has a simultaneous feedback of these waves on the electrons
\citep{vvs}. This mechanism is described by QLD, leading to the
diffusion of particles as along as across the magnetic field lines.
The process of QLD in the external magnetic field is examined in a
series of books \citep{melr,axiez}. Generally speaking, at the
pulsar surface relativistic particles efficiently loose their
perpendicular momenta via synchrotron emission in very strong
($B\sim10^{12}$G) magnetic fields and therefore, they very rapidly
transit to their ground Landau state (pitch angles are vanishing).
Contrary to this process, QLD leads to creation of the pitch angles
by resonant particles and as a result they start to radiate in the
synchrotron regime. To explain the observed very high energy
emission of the Crab pulsar, it is supposed that the resonant
particles are the primary beam electrons with
$\gamma_{b}\sim10^{8}$, giving the synchrotron emission in the
VHE domain.

When emitting in the synchrotron regime, the resonant particles
undergo the radiation reaction force ${\bf F}$, having as
longitudinal as transversal components \citep{landau}:
\begin{equation}\label{4}
    F_{\perp}=-\alpha_{s}\frac{p_{\perp}}{p_{\parallel}}\left(1+\frac{p_{\perp}^{2}}{m^{2}c^{2}}\right),
    F_{\parallel}=-\frac{\alpha_{s}}{m^{2}c^{2}}p_{\perp}^{2},
\end{equation}
where $\alpha_{s}=2e^{2}\omega_{B}^{2}/3c^{2}$.

The wave excitation leads to a redistribution process of the
particles via QLD. The kinetic equation for the distribution
function of the resonant particles can be written as
\citep{machus1,malov02}:
\begin{eqnarray}\label{5}
\frac{\partial\textit{f }^{0}}{\partial
    t}+\frac{\partial}{\partial
p_{\parallel}}\left\{F_{\parallel}\textit{f
}^{0}\right\}+\frac{1}{p_{\perp}}\frac{\partial}{\partial
p_{\perp}}\left\{p_{\perp}F_{\perp}\textit{f }^{0}\right\}=\nonumber
\\=\frac{1}{p_{\perp}}\frac{\partial}{\partial p_{\perp}}\left\{p_{\perp}\left(D_{\perp,\perp}\frac{\partial}{\partial p_{\perp}}+D_{\perp,\parallel}\frac{\partial}{\partial
p_{\parallel}}\right)\textit{f
}^{0}\left(\mathbf{p}\right)\right\}+\nonumber
 \\
+\frac{\partial}{\partial
p_{\parallel}}\left\{\left(D_{\parallel,\perp}\frac{\partial}{\partial
p_{\perp}}+D_{\parallel,\parallel}\frac{\partial}{\partial
p_{\parallel}}\right)\textit{f }^{0}\left(\mathbf{p}\right)\right\}.
\end{eqnarray}

The diffusion coefficients in Eq. (5) are evaluated in the momentum
space as \citep{melr}:
\begin{eqnarray}\label{6}
    \left(%
\begin{array}{c}
  D_{\perp,\perp} \\
  D_{\perp, \parallel}=D_{\parallel\perp} \\
  D_{\parallel,\parallel} \\
\end{array}%
\right)= \int \frac{d^{3}\mathbf{k}}{(2\pi)^{3}}\frac{
\pi^{2}e^{2}\upsilon^{2}\sin^{2}\psi}{\hbar^{2}\omega^{2}}\frac{|E_{k}|^{2}}{4\pi}\times\nonumber\\
\times\delta(\omega(\mathbf{k})-k_{\parallel}\upsilon_{\parallel}+\omega_{B}/\gamma)
\left(%
\begin{array}{c}
  (\Delta p_{\perp})^2 \\
  (\Delta p_{\perp})(\Delta p_{\parallel})  \\
  (\Delta p_{\parallel})^2 \\
\end{array}%
\right).
\end{eqnarray}

Where $|E_{k}|^{2}/4\pi$ is the density of electric energy in the
excited waves and
\begin{equation}\label{7}
    \Delta p_{\perp}=-\frac{\hbar\omega_{B}}{\gamma \upsilon \sin \psi}, \qquad
    \Delta p_{\parallel}=\hbar k_{\parallel}.
\end{equation}
The evaluation in our case gives
\begin{eqnarray}\label{8}
     \left(%
\begin{array}{c}
  D_{\perp,\perp} \\
  D_{\perp, \parallel}=D_{\parallel\perp} \\
  D_{\parallel,\parallel} \\
\end{array}%
\right)=\left(%
\begin{array}{c}
  D\delta|E_{k}|^{2}_{k=k_{res}} \\
  -D \psi|E_{k}|^{2}_{k=k_{res}}  \\
  D \psi^{2}\frac{1}{\delta}|E_{k}|^{2}_{k=k_{res}} \\
\end{array}%
\right),
\end{eqnarray}
where $D=e^{2}/8c$.

The pitch-angle acquired by resonant particles during the process of
QLD satisfies $\psi= p_{\perp}/p_{\parallel}\ll1$. Thus, one can
assume $\partial/\partial p_{\perp}>>\partial/\partial
p_{\parallel}$ which reduces Eq. (5) to the following form

\begin{eqnarray} \label{9}
    \frac{\partial\textit{f }^{0}}{\partial
    t}+\frac{1}{p_{\perp}}\frac{\partial}{\partial p_{\perp}}\left(p_{\perp}
    F_{\perp}\textit{f }^{0}\right)=\nonumber \\
    =\frac{1}{p_{\perp}}\frac{\partial}{\partial p_{\perp}}\left(p_{\perp}
D_{\perp,\perp}\frac{\partial\textit{f }^{0}}{\partial
p_{\perp}}\right).
\end{eqnarray}
The transversal diffusion leads to the izotropization of the
one-dimensional distribution function whereas the force $F_{\perp}$
works against the diffusion. The dynamical process saturates when
these effects balance each other. Considering the
quasi-stationary state ($\partial\textit{f}/\partial t=0$), one
finds
\begin{equation}\label{10}
    \textit{f}(p_{\perp})=C exp\left(\int \frac{F_{\perp}}{D_{\perp,\perp}}dp_{\perp}\right)=Ce^{-\left(\frac{p_{\perp}}{p_{\perp_{0}}}\right)^{4}},
\end{equation}
where
\begin{equation}\label{11}
     p_{\perp_{0}}\approx\frac{\pi^{1/2}}{B\gamma_{p}^{2}}\left(\frac{3m^{9}c^{11}\gamma_{b}^{5}}{32e^{6}P^{3}}\right)^{1/4}.
\end{equation}
Taking into account the above relation we can find that the mean
value of the pitch-angle, $\psi_0\approx
p_{\perp_{0}}/p_{\parallel}$, is of the order of $10^{-6}$.

\section{Synchrotron radiation spectrum}

Let us consider the synchrotron emission of the set of electrons. If
$p_{\perp}\textit{f }d p_{\perp}dp_{\parallel}dVd\Omega_{\tau}$ is
the number of emitting particles in the elementary $dV$ volume, with
momenta from the intervals $[p_{\perp}, p_{\perp}+dp_{\perp}]$ and
$[p_{\parallel},p_{\parallel}+dp_{\parallel}]$, and with the
velocities that lie inside the solid angle $d\Omega_{\tau}$ near the
direction of $\vec{\tau}$. Then the emission flux of the set of
electrons is given by \citep{ginz}
\begin{equation}\label{12}
    F_{\epsilon}=\int I_{e}p_{\perp}\textit{f }d p_{\perp}dp_{\parallel}dVd\Omega_{\tau},
\end{equation}
where $I_{e}$ is the Stokes parameter, which is additive in this
case, as the observed synchrotron radiation wavelength $\lambda$ is
much less than the value of $n^{-1/3}$ - the average distance
between particles, where $n$ is the density of plasma component
electrons. Taking into account that
\begin{equation}\label{13}
    \int p_{\perp}
    \textit{f }^{0}dp_{\perp}=\textit{f}_{\parallel}(p_{\parallel}),
\end{equation}
the integral (\ref{16}) is easily reduced to
\begin{equation}\label{14}
    F_{\epsilon}\propto\int\textit{f}_{\parallel}(p_{\parallel})B\psi\frac{\epsilon}{\epsilon_{m}}\left[\int_{\epsilon/
    \epsilon_{m}}^{\infty}K_{5/3}(z)dz \right] dp_{\parallel}.
\end{equation}
Here $\epsilon_{m}\approx5\cdot10^{-18}B\psi\gamma^{2}$GeV is the
photon energy of the maximum of synchrotron spectrum of a single
electron and $K_{5/3}(z)$ is a Macdonald function. After
substituting the mean value of the pitch-angle in the above
expression for $\epsilon_{m}$, we get
\begin{equation}\label{15}
    \epsilon_{m}\simeq5\cdot10^{-18}\frac{\pi^{1/2}}{\gamma_{p}^{2}}\left(\frac{3m^{5}c^{7}\gamma_{b}^{9}}{4e^{6}P^{3}}\right)^{1/4}
\end{equation}
Accordingly, the beam electrons should have
$\gamma_{b}\simeq6\cdot10^{8}$ to radiate the photons with
$\sim10$GeV energy. This in turn implies that the gap
models providing the Lorentz factors $\sim 10^7$, are not enough to
explain the detected pulsed emission. On the other
hand, \cite{magic} confirmed that their observations indicate that
emission happens far out in the
magnetosphere. One of the
real scenarios could be the centrifugal acceleration of electrons,
which take place in co-rotating magnetospheres
\citep{mr,r03,osm7}. Another alternative mechanism of acceleration
could be a collapse
\citep{arcim,zax} of the centrifugally excited unstable Langmuir
waves \citep{incr1} in the pulsar's magnetosphere.

According to our emission model, the observed radiation comes from a
region where the magnetic field lines are practically straight and
parallel to each other, therefore, electrons with $\psi\approx\psi_0$
efficiently emit in the observer's direction.

To find the synchrotron flux in our case, we need to know the
one-dimensional distribution function of the emitting particles $
\textit{f}_{\parallel}$. Let us multiply both sides of Eq. (5) on
$p_{\perp}$ and integrate it over $p_{\perp}$. Taking into account
that the distribution function vanishes at the boundaries of
integration, Eq. (\ref{5}) reduces to
\begin{eqnarray}\label{16}
    \frac{\partial\textit{f}_{\parallel}}{\partial t}=\frac{\partial}{\partial
    p_{\parallel}}\left({\frac{\alpha_{s}}{m^{2}c^{2}\pi^{1/2}}p_{\perp_{0}}^{2}\textit{f}_{\parallel}}\right).
\end{eqnarray}
Considering the quasi-stationary case we find
\begin{eqnarray}\label{17}
    \textit{f}_{\parallel}\propto\frac{1}{p_{\parallel}^{1/2}|E_{k}|}.
\end{eqnarray}

For $\gamma\psi\ll10^{10}$, a magnetic field inhomogeneity does not
affect the process of wave excitation. The equation that describes
the cyclotron noise level, in this case, has the form \citep{lomi}
\begin{equation}\label{18}
    \frac{\partial|E_{k}|^{2}}{\partial
    t}=2\Gamma_{c}|E_{k}|^{2}\textit{f}_{\parallel},
\end{equation}
where
\begin{equation}\label{19}
   \Gamma_{c}=\frac{\pi^{2}e^{2}}{k_{\parallel}}\textit{f}_{\parallel}(p_{res}),
\end{equation}
is the growth rate of the instability. Here $k_{\parallel}$ can be
found from the resonance condition (3)
\begin{equation}\label{20}
     k_{\parallel_{res}}\approx\frac{\omega_{B}}{c\delta\gamma_{res}}.
\end{equation}
Combining Eqs. (16) and (18) one finds
\begin{equation}\label{21}
    \frac{\partial }{\partial t}\left\{\textit{f}_{\parallel}-\alpha\frac{\partial}{\partial
    p_{\parallel}}\left(\frac{|E_{k}|}{p_{\parallel}^{1/2}}\right)\right\}=0,
\end{equation}
\begin{equation}\label{22}
    \alpha=\left(\frac{4}{3}\frac{e^{2}}{\pi^{5}c^{5}}\frac{\omega_{B}^{6}\gamma_{p}^{3}}{\omega_{p}^{2}}\right)^{1/4},
\end{equation}
which reduces to
\begin{equation}\label{23}
    \left\{\textit{f}_{\parallel}-\alpha\frac{\partial}{\partial
    p_{\parallel}}\left(\frac{|E_{k}|}{p_{\parallel}^{1/2}}\right)\right\}=const.
\end{equation}

Taking into account that for the initial moment the major
contribution of the lefthand side of the Eq. (\ref{23}) comes from
$\textit{f}_{\parallel_{0}}$, the corresponding expression writes as
\begin{equation}\label{24}
   \textit{f}_{\parallel}-\alpha\frac{\partial}{\partial
    p_{\parallel}}\left(\frac{|E_{k}|}{p_{\parallel}^{1/2}}\right)=\textit{f}_{\parallel_{0}}.
\end{equation}
The distribution function $\textit{f}$ is proportional to
$n\sim1/r^{3}$ (here $r$ is the distance from the pulsar), then one
should neglect $ \textit{f}_{\parallel}$ in comparison with
$\textit{f}_{\parallel_{0}}$. Consequently, the above equation
reduces to
\begin{equation}\label{25}
   \alpha\frac{\partial}{\partial
    p_{\parallel}}\left(\frac{|E_{k}|}{p_{\parallel}^{1/2}}\right)+\textit{f}_{\parallel_{0}}=0.
\end{equation}
As we can see the function $E_{k}(p_{\parallel})$ drastically
depends on the form of the initial distribution of the primary beam
electrons. According to the work, \citep{gold69}, a spinning
magnetized neutron star generates an electric field which extracts
electrons from the star's surface and accelerates them to form a
low-density ($n_{b}=B/Pce$) and energetic primary beam. We only know
the scenario of creation of the primary beam, but nothing can be
told about its distribution, which drastically depends on the
neutron star surface properties and temperature. To our knowledge
there is no convincing theory which could predict the form of the
distribution function of the beam electrons. Thus, we can only
assume that the beam electrons have a power-law distribution
\begin{equation}\label{26}
     \textit{f}_{\parallel_{0}}\propto p_{\parallel}^{-n},
\end{equation}
and for the energy density of the waves we get
\begin{equation}\label{27}
     |E_{k}|^{2}\propto p_{\parallel}^{3-2n}.
\end{equation}
The effective value of the pitch angle depends on $|E_{k}|^{2}$ as
follows
\begin{equation}\label{28}
    \psi_{0}=\frac{1}{2\omega_{B}}\left(\frac{3m^{2}c^{3}}{p_{\parallel}^{3}}\frac{\omega_{p}^{2}}{\gamma_{p}^{3}}|E_{k}|^{2}\right)^{1/4}.
\end{equation}
Using expression (17), (27) and (28), and replacing the integration
variable $p_{\parallel}$ by $x=\epsilon/\epsilon_{m}$, from Eq.
(\ref{14}) we will get
\begin{equation}\label{29}
    F_{\epsilon}\propto\epsilon^{-\frac{2-n}{4-n}}\int x^{\frac{2-n}{4-n}}\left[\int_{x}^{\infty}K_{5/3}(z)dz \right] dx.
\end{equation}
According to \citet{magic} the observed high energy pulsed emission
of the crab pulsar is best described by a power-law spectrum
$F(\epsilon)\propto\epsilon^{-2.022}$ in the energy domain
$(0.01-5)$GeV. At $\epsilon=25$GeV a measured flux is several times
lower, which requires a spectral cutoff somewhere between $5$ and
$25$GeV.

We assume that the energy of the beam electrons vary between
$\gamma_{min}\sim10^{6}$ and $\gamma_{max}\sim10^{8}$, in which
case, we have $(\epsilon/\epsilon_{m})_{max}\ll1$ and
$(\epsilon/\epsilon_{m})_{min}\gg1$. Under such conditions the
integral (29) can be approximately expressed by the following
function
\begin{equation}\label{30}
    F_{\epsilon}\propto\epsilon^{-\frac{2-n}{4-n}}exp\left[-\left(\frac{\epsilon}{23}\right)^{1.6}\right].
\end{equation}
When $n=6$ the spectral index, $\beta$, of the synchrotron emission equals
$2$, and the flux
$F_{\epsilon}\propto\epsilon^{-2}exp[-(\epsilon/23)^{1.6}]$. As we
can see our emission scenario predicts the exponential cutoff, with
the cutoff energy $23$GeV.

\section{Discussion}

One of the interesting observational feature of the Crab pulsar is
that its multiwavelength emission pulses from low-frequency radio
waves up to hard $\gamma$-rays ($\epsilon>25$GeV) are coincident in
phase \citep{manch,magic}. Which implies that generation of these
waves occurs in the same place of the pulsar magnetosphere.
According to the generally accepted point of view, VHE emission is
produced either by the Inverse Compton up-scattering or by the
curvature radiation. Although it is clear that the aforementioned
processes cannot provide the observationally evident coincidence of
signals, since they do not have any restriction on the spacial
location of emission (area in the pulsar magnetosphere, where the
corresponding radiation is produced). This particular problem has
been studied by \citet{difus1}. Considering the curvature radiation,
we have shown that the curvature drift instability
(\cite{mnras,forcefree}) makes the magnetic field lines rectify very
efficiently. It has been shown that the increment of the instability
is given by [see Eq. 22 in (\cite{forcefree})]
\begin{equation}
\label{increm} \Gamma \approx
\left(-\frac{3}{2}\frac{\omega^2_{b}}{\gamma_{b_0}}\frac{k_xu_{x}}{k_{r}c}\right)
^{1/2}\left|J_0\left(\frac{k_xu_{x}}
{4\Omega}\right)J_{0}\left(\frac{k_{r}c}{\Omega}\right)\right|,
\end{equation}
where $\omega_{b}$ is the plasma frequency of the beam component,
$k_r$ and $k_x$ are the wave vector's radial component and the
component along the rotation axis respectively and $J_0$ denotes the
Bessel function of zeroth order. By considering the perturbation
corresponding to $\lambda_x\sim R_{lc}$, $\lambda_r\sim 10^3R_{lc}$
and the initial curvature of the magnetic field lines being of the
order of $R_{lc}$, where $R_{lc}\sim 10^8$cm is the light cylinder
lengthscale, one can show that the timescale, $t_{CDI}\sim
1/\Gamma$, of the CDI for $\gamma_b\sim 10^8$ equals $1.6$s. On the
other hand, the instability makes its job (amplifies the toroidal
magnetic field) until the excited mode escapes the magnetosphere.
This happens in the characteristic timescale $t_{esc}\sim
R_{lc}/(\upsilon_{ph}\sin\theta)$, where
$\upsilon_{ph}\equiv\omega/k$ is the phase velocity of the curvature
drift wave and $\theta\approx k_r/k_x$ is the inclination of the
wave vector with respect to the rotation axis. After taking into
account the dispersion relation of the curvature drift mode,
$\omega=k_xu_x/2$ (\cite{mnras,forcefree}), it is straightforward to
show that $t_{esc}\approx 2.5\times 10^{3}$s. Therefore, the
timescales satisfy the condition $t_{CDI}\ll t_{esc}$, implying that
the curvature drift instability is efficient enough to rectify the
magnetic field lines (curvature tends to zero), leading to a
negligible role of the curvature emission process in the observed
VHE domain. It is worth noting that the wave stays in the active
zone longer than the plasma which caused this wave. On the other
hand, the CDI is a continuous process in spite of the fact that the
mode escapes the magnetosphere, because a new portion of plasma
excites the CDI again and the process is continuously maintained.

By analyzing the inverse Compton scattering, we have found that for
Crab pulsar's magnetospheric parameters even very energetic
electrons are unable to produce the observed photon energies.


The emission model proposed in previous  works \cite{difus,difus1}
and developed in the present paper ensures the simultaneous
generation of the low and high frequency waves in the same area of
the magnetosphere. The distribution function of relativistic
particles is one dimensional at the pulsar surface, but plasma with
an anisotropic distribution function is unstable which inevitably
leads to the wave excitation. The main mechanism of the wave
generation in plasmas of the pulsar magnetosphere is the cyclotron
instability, which develops near the light cylinder. During the
quasi-linear stage of the instability a diffusion of particles
arises along and across the magnetic field lines. Therefore, plasma
particles acquire transverse momenta and, as a result, the
synchrotron mechanism is switched on. If the resonant particles are
the primary beam electrons with $\gamma_{b}\simeq6\cdot10^{8}$ their
synchrotron emission comes in the high energy domain ($\sim 10$GeV).
The frequency of the original waves, excited during the cyclotron
resonance can be estimated from Eq. (3) as follows
$\omega_{0}\approx\omega_{B}/\delta\gamma_{b}$. Estimations show
that for the beam electrons with the Lorentz-factor from the
interval $\gamma_{b}\sim 10^{6-8}$, the radio waves are excited.
Consequently, we explain the coincidence of radio and $\gamma$-ray
signals.

We provide the theoretical confirmation of the measured power-law
spectrum ($F_{\epsilon}\propto\epsilon^{-\beta}$ with $\beta=2$) in
the energy domain $\epsilon=0.01$GeV to $25$Gev. Differently from
the standard theory of the synchrotron emission \citep{ginz}, which
only explains the spectral index, $\beta<1$, our approach gives the
possibility to obtain the values of index that are much higher than
one. The main reason for this is that we take into account the
mechanism of creation of the pitch angles, and obtain a certain
distribution function of the emitting particles from their
perpendicular momenta (see Eq. (10)), which restricts the possible
values of the pitch angles. The emission comes from a region of the
pulsar magnetosphere where the magnetic field lines are practically
straight and parallel to each other. But in the standard theory of
the synchrotron emission \citep{ginz}, it is supposed that the
observed radiation is collected from a large spacial region in
various parts of which, the magnetic field is oriented randomly.
Thus, it is supposed that along the line of sight the magnetic field
directions  are chaotic and when finding emission flux, Eq. (14) is
averaged over all directions of the magnetic field (which means
integration over $\psi$ varying from $0$ to $\pi$). The measured
decrease of the flux at $\epsilon=25$GeV is also explained. Our
theoretical spectrum
$F_{\epsilon}\propto\epsilon^{-2}exp[-(\epsilon/23)^{1.6}]$ yields
the exponential cutoff, with the cutoff energy $23$GeV.

\section{Summary}\label{sec:summary}

\begin{enumerate}

      \item Constructing a self-consistent theory,
we interpret the observations of the MAGIC Cherenkov Telescope of
the pulsed emission, $(0.01-25)$GeV, from the Crab pulsar.

      \item It is emphasized that due to very
small cooling timescales, particles rapidly transit to the ground
Landau state vanishing the subsequent synchrotron radiation.
The situation changes thanks to the cyclotron
instability, which  efficiently develops on the light cylinder scales
and creates non-vanishing pitch angles, leading to the efficient
synchrotron process.

\item The observational fact of the coincidence
of signals in low (radio) and VHE domains is
explained. The original waves excited during the cyclotron
instability come in the radio band. It is shown that the resonant electrons
interact with the aforementioned waves via QLD, acquire
pitch angles and start to radiate in the synchrotron regime.

\item Considering a new approach of the synchrotron theory based
on our emission model,
we have found the spectral index, $\beta$, of VHE emission to be equal to
$2$ and the exponential cutoff, with the cutoff
energy - $23$GeV, being in a good agreement with the observational
data \citep{magic}.

      \end{enumerate}
In the present paper, we have shown the spacial coincidence of the
region of generation of radio and VHE emission of the Crab pulsar.
Although, the aforementioned coincidence of signals is detected for
broad frequency ranges (from radio up to hard $\gamma$-rays). Thus,
we suppose that the generation of the multiwavelength radiation of
the Crab pulsar takes place in one location of the pulsar
magnetosphere. In particular, if the cyclotron resonance occurs for
the tail electrons, the above described processes might cause the
simultaneous generation of waves in a different energy domains, which is a
topic of our future work.

\section*{Acknowledgments}
The research was supported by the Georgian National Science
Foundation grant GNSF/ST06/4-193.



\begin{thebibliography}{}
\bibitem[Akhiezer(1967)]{axiez} Akhiezer A.I., 1967,
Collective oscillations in a plasma, M.I.T. Press (1967)
\bibitem[Aliu et al.(2008)]{magic} Aliu E.
et al., 2008, Sci, 322, 1221A
\bibitem[Arons(1981)]{arons} Arons J., 1981, in Proc. Varenna Summer School
and Workhop on Plasma Astrophysics, ESA, 273
\bibitem[e.g. Artsimovich \& Sagdeev(1979)]{arcim} Artsimovich L.A. \& Sagdeev R.Z.,
1979, Plasma physics for physicists, (russian edition) Moscow,
Atomizdat
\bibitem[Bekefi \& Barrett (1977)]{bekefi} Bekefi George \& Barrett Alan H., 1977,
Electromagnetic vibrations, waves and radiation, The MIT Press,
Cambridge, Massachusetts and London, England
\bibitem[Daugherty \& Harding (1982)] {dh} Daugherty, J. K., \&
Harding, A. K., 1982, \apj, 252, 337
\bibitem[Dyks et al. (2004)] {dyks} Dyks, J., Rudak, B., Harding, A.
K., 2004, \apj, 607, 939
\bibitem[Ginzburg(1981)]{ginz} Ginzburg, V.L., 1981, "Teor.
Fizika i Astrofizika", Nauka M. 1981

\bibitem [Goldreich \& Julian (1969)] {gold69} Goldreich, P., Julian, W. H., 1969, ApJ, 157, 869
\bibitem[Kazbegi et al.(1992)]{kmm} Kazbegi A.Z., Machabeli G.Z
\& Melikidze G.I., 1992, in Proc. IAU Collog. 128, The
Magnetospheric Structure and Emission Mechanisms of Radio Pulsars,
ed. T.H. Hankins, J.M. Rankin \& J.A. Gil (Zielona Gora: Pedagogical
Univ. Press), 232
\bibitem[Kuiper et al.(2001)]{kuiper} Kuiper L., Hermsen W., Cusumano G., Diehl R., Schönfelder
\bibitem[Landau \& Lifshitz(1971)]{landau} Landau L.D. \& Lifshitz E.M.,
51971, Classical Theory of Fields (London: Pergamon)
\bibitem[Lyutikov et al.(1999)]{lmb} Lyutikov M., Machabeli
G. \& OBlandford R., 1999, \apj, 512, 804
\bibitem[Lominadze et al.(1979)]{lomin}
Lominadze J.G., Machabeli G.Z. \& Mikhailovsky A.B., 1979, J. Phys.
Colloq., 40, No. C-7, 713
\bibitem[Lominadze et al. (1983)] {lomi} Lominadze J.G., Machabeli
G. Z., \& Usov V. V., 1983, Ap\&SS, 90, 19L
\bibitem[Machabeli \& Usov(1979)]{machus1} Machabeli G.Z. \& Usov
V.V., 1979, AZhh Pis'ma, 5, 445
\bibitem[Machabeli et al.(2005)]{incr1} Machabeli G., Osmanov
Z. \& Mahajan S., 2005, Phys. Plasmas 12, 062901
\bibitem[Machabeli \& Osmanov(2009)]{difus} Machabeli G. \& Osmanov
Z., 2009, \apjl, 700, 114
\bibitem[Machabeli \& Osmanov(2010)]{difus1} Machabeli G. \& Osmanov
Z., 2010, \apj, 709, 547
\bibitem[Machabeli \& Rogava(1994)]{mr} Machabeli, G.Z. \&
Rogava, A. D., 1994, Phys.Rev. A, 50, 98

\bibitem [Malov \& Machabeli (2002)]{malov02} Malov, I, F.,
Machabeli, G. Z., 2002, Astronomy Reports, Vol. 46, Issue 8, p.684
\bibitem[Manchester \& Taylor(1980)]{manch} Manchester R.N. \&
Taylor J.H., 1980, Pulsars, F.H. Freeman and Company

\bibitem[Melrose \& McPhedran(1991)]{melr} Melrose D.B. \& McPhedran R.C., 1991,
Electromagnetic Processes in Dispersive Media, Cambridge University
Press (September 27, 1991)
\bibitem[Morini (1983)] {mo} Morini, M., 1983, MNRAS, 202, 495

\bibitem[see Osmanov et al.(2008)]{mnras} Osmanov, Z., Dalakishvili, Z. \& Machabeli, Z., 2008, \mnras, 383, 1007
\bibitem[Osmanov et al.(2007)]{osm7} Osmanov Z., Rogava A.S. \& Bodo G., 2007, \aap, 470,
395
\bibitem[Osmanov et al.(2009)]{forcefree} Osmanov, Z., Shapakidze, D.
\& Machabeli, Z. 2009, \aap, 503, 19
\bibitem[Rogava et al.(2003)]{r03} Rogava A. D., Dalakishvili G. \& Osmanov Z., 2003, Gen. Relativ. Gravit. 35, 1133
\bibitem[Romani \& Yadigaroglu (1995)] {RY} Romani, R. W., \&
Yadigaroglu, I. A., 1995, \apj, 438, 314
\bibitem[Shapakidze et al.(2003)]{smmk} Shapakidze D., Machabeli G., Melikidze G. \& Khechinashvili D., 2003, Phys. Rev. E., id. 026407
\bibitem[Sturrock(1971)]{stur} Sturrock P.A., 1971, \apj, {\bf 164}, 529
\bibitem[Tademaru(1973)]{tadem} Tademaru E., 1973,
\apj, {\bf 183}, 625


\bibitem[Vedenov et al.(1961)]{vvs} Vedenov A.A., Velikhov E.P. \& Sagdeev
R.Z., 1961, Soviet Physics Uspekhi, Volume 4, Issue 2, 332
\bibitem[Zakharov(1972)]{zax} Zakharov V.E., 1972, JETP, {\bf 35},
908


\end{thebibliography}
\end{document}